# Introducing the treatment hierarchy question in network meta-analysis

Georgia Salanti[1], Adriani Nikolakopoulou[1,2], Orestis Efthimou[1], Dimitris Mavridis[3], Matthias Egger[1,4], Ian R. White[5]

[1] Institute of Social and Preventive Medicine, University of Bern, Switzerland

[2] Institute of Medical Biometry and Statistics, Faculty of Medicine and Medical Center - University of Freiburg, Germany

[3]Department of Primary Education, University of Ioannina, Greece

[4] Population Health Sciences, Bristol Medical School, University of Bristol, Bristol, UK

[5] MRC Clinical Trials Unit at UCL, UK

**ACKNOWLEDGEMENTS**

This work is supported by a Swiss National Science Foundation grant No 179158 (GS, AN).

IW was supported by the Medical Research Council Programme MC_UU_12023/21.

OE is supported by the Swiss National Science Foundation, grant No 180083.

**Abstract**

*Background:* Comparative effectiveness research using network meta-analysis can present a hierarchy of competing treatments, from the least to most preferable option. However, the research question associated with the hierarchy of multiple interventions is never clearly defined in published reviews.
*Methods and Results:* We introduce the notion of a *treatment hierarchy question* that describes the criterion for choosing a specific treatment over one or more competing alternatives. For example, stakeholders might ask which treatment is most likely to improve mean survival by at least 2 years or which treatment is associated with the longest mean survival. The answers to these two questions are not necessarily the same. We discuss the most commonly used ranking metrics (quantities that describe or compare the estimated treatment-specific effects), how the metrics produce a treatment hierarchy and the type of treatment hierarchy question that each metric can answer. We show that the ranking metrics encompass the uncertainty in the estimation of the treatment effects in different ways, which results in different treatment hierarchies.
*Conclusions:* Network meta-analyses that aim to rank treatments should state in the protocol the *treatment hierarchy question* they aim to address and employ the appropriate ranking metric to answer it.

In comparative effectiveness research, the ranking of multiple competing treatments is a potentially important advantage of network meta-analysis (NMA) over pairwise meta-analysis (1–3). Different methods have been proposed in the literature, including estimation of the probability of each treatment assuming each rank, the median and mean rank of each treatment and the surface under the cumulative ranking curve (SUCRA) for each treatment. SUCRA is a (Bayesian) summary of the rank distribution, which can be interpreted as the estimated proportion of treatments worse than the treatment of interest. It can be approximated by a frequentist analogue, the P-score (2,4).

Several tutorials correctly point out that while ranking metrics may be useful, the relative treatment effects and their uncertainty are the most clinically relevant output from NMA (5–7). Criticism of the use of ranking metrics is abundant in the literature. For example, Kibret et al. performed a simulation study and concluded that «decisions should not be made based on rank probabilities especially when treatments are not directly compared as they may be ill-informed» (8). Mills et al. concluded that «interpretability [of treatment ranks] is limited by the fact that they are driven predominantly by the estimated effect sizes, and that standard errors play an unduly small role in determining their position» (9). Veroniki et al. and Trinquart et al. examined published several networks and concluded that «the ranking statistic values might be unstable» (10,11). Wang and Carter stated that «SUCRA findings can be misleading and should be interpreted with caution» (12).

These concerns have raised awareness of what ranking can and cannot do and drawn attention to the dangers of oversimplification and reliance on treatment hierarchy alone. However, some of the criticisms inappropriately attribute the problem to the ranking metrics per se. In this educational article we argue that ranking metrics are not misleading. Rather, the confusion in the literature is due to the fact that different ranking metrics aim to answer different questions about treatment hierarchy, and that researchers do not clearly define what they mean by "the best treatment" in a given setting. A review of 232 NMAs suggested that none of them reported the definition of the preferable treatment, or anything that could be interpreted as what we later define as a *'treatment hierarchy question' (13)*. We show that this confusion is specific to networks and does not arise in comparisons of two treatments. We also suggest that any comparison of the treatment hierarchies obtained by different ranking metrics should acknowledge that they provide answers to different treatment hierarchy questions.

We first introduce the notion of the treatment hierarchy question and argue that the question addressed by an NMA should determine the quantities used to answer this question (i.e. the ranking metrics used to obtain the treatment hierarchy). We then present the properties of the commonly used ranking metrics and discuss the questions they address. We use theoretical examples to show that each ranking metric encompasses the uncertainty in the estimation of the relative treatment effects in a different way, and this can lead to different treatment hierarchies. We conclude with recommendations and a discussion about possible future extensions of the existing ranking metrics, which can answer more complex treatment hierarchy questions.

All results, figures and tables are produced in R software and are reproducible with the scripts in the GitHub repository esm-ispm-unibe-ch-REPRODUCIBLE/RankingPaper.

## WHAT IS A TREATMENT HIERARCHY QUESTION?

### Defining a treatment hierarchy question

Careful and clear framing of the research question before starting any systematic review is good practice, to which most published NMAs adhere. In a recent bibliographic study, all published NMAs clearly stated the efficacy and safety outcomes, and the effect measure(s) used to compare each pair of competing interventions (14). The focus of evidence synthesis in comparative effectiveness research is the relative treatment effect: for example, the difference in the mean of a quantitative outcome between treatments. The clinical trials literature now calls this quantity the *estimand*: the quantity that is to be estimated (15). In this paper, we consider *absolute* estimands such as the treatment-specific mean as well as *relative* estimands which are comparisons of absolute estimands. In the case when two treatments are compared, if we knew the true value of the relative treatment effect, this would automatically define which treatment is preferable; in practice, the relative treatment effect is estimated with error, but its value still shows which treatment appears preferable.

Similar attention to a clear definition of the research question has not been paid when several treatments are compared. Authors do not report a clear definition of what they mean by the 'best' or 'preferable' treatment. In NMA we have several relative treatment effects. If they were known without error, then the treatment hierarchy would be clear. When the relative treatment effects are estimated with error, however, they cannot easily be translated into a treatment hierarchy. As discussed later in this article, the crucial difficulty arises when we have to deal with uncertainty around multiple relative treatment effects.

A *treatment hierarchy question* describes the criterion used to prefer one treatment over another or several competing treatments. The question must be answerable from the available data and must relate to the selected estimand. We take the example of treatments to lower low-density lipoprotein cholesterol (LDL-C). "Which is the best treatment?" is not an appropriate treatment hierarchy question because it cannot be answered with certainty from data. Instead, we can answer the question "Which treatment is most likely to be the best treatment?" by using data to make probability statements about the treatment effects. Next, the term "best" needs to be defined by linking it to an estimand. Such refinement of the question might, for example, lead to the clear question "Which treatment is most likely to produce a mean LDL-C level of ≤ 2.5 mmol/L?". In this case, we will need to use our data to calculate the probabilities for each drug that the mean LDL-C does not exceed 2.5 mmol/L and then order the drugs according to those probabilities.

The criterion used to identify the best treatment often aims to maximize a summary statistic or estimate of the (beneficial) impact of the treatment on one or more health outcomes. We call a *ranking metric* any quantity that describes the impact of a treatment (relative or not to other treatments). In the question above, the relevant ranking metric is the probability that the mean LDL-C is ≤ 2.5 mmol/L. Then the answer to the treatment hierarchy question is given by maximizing this probability across all treatment options, and we call this the preferable treatment. Further examples of ranking metrics are discussed below.

**Setting and notation**

Consider several drugs denoted by $i$ ($i = 1, ..., T$) and a single harmful outcome of interest, where the estimands are the (absolute) true means of the outcome $\mu_i$ or their (relative) differences $\delta_{ij} = \mu_i - \mu_j$. Because we have finite sample sizes, our estimates of $\mu_i$ and $\delta_{ij}$ come with uncertainty. In what follows, any reference to the estimation of $\mu_i$ or $\delta_{ij}$ will involve a whole distribution of possible values, and *ranking metrics* will describe features of these distributions (one for each $i$). One popular way to communicate the uncertainty in each distribution is to present a range of plausible values such as a 95% confidence interval, or if a Bayesian approach is taken, a 95% credible interval. At the center of the distribution that estimates $\mu_i$ is the point estimate $M_i$, our 'most likely' estimate of the true mean outcome with treatment $i$. $M_i$ is a single known value (unlike the unknown $\mu_i$), and it is one of the many possible *ranking metrics* discussed below. Similarly, the point estimate of the relative treatment effect $\delta_{ij}$ is $D_{ij} = M_i - M_j$. We consider that a treatment $i$ 'beats' treatment $j$ when $\mu_i < \mu_j$; we can use the distribution that estimates $\mu_i$ and $\mu_j$ to estimate the probability that treatment $i$ beats treatment $j$. A summary of the notation is presented in Table 1.

**Example: formulating treatment hierarchy questions for interventions to reduce LDL-C levels**

Consider the fictional example of three drugs aiming to lower LDL-C levels in patients at high risk of cardiovascular disease. Suppose for a moment that only LDL-C levels determine the preferable treatment, though in reality we should also consider the effects of the drug on HDL, cardiovascular events and mortality and also cost and convenience. After synthesis of RCTs that compare pairs of cholesterol lowering drugs in participants with baseline LDL-C levels between 2.60 and 5.10 mmol/L, suppose that our knowledge about the true population mean post-treatment levels $\mu_A, \mu_B, \mu_C$ is shown in the distributions of Figure **1**. These distributions are characterized by the three centers (or point estimates) $M_A, M_B, M_C$ and uncertainty that depends on the amount of information available for each treatment, with C having the most information and B the least.

There are several treatment hierarchy questions one can ask and the order of the treatments depends on that question. One possible question is

*"Which treatment has the smallest estimated mean post-treatment LDL-C level?"*

(treatment hierarchy question 1).

This orders the treatments according to $M_i$ and indicates B as the preferable treatment.

Alternatively, we can heuristically interpret the European guidelines, that recommend that treatment should halve LDL-C levels, to obtain another treatment hierarchy question (16). In a population with an average 5 $mmol/L$, we can set the goal to have post-treatment mean value $\mu_i$ below 2.5 mmol/L, and a possible question is then

*"Which treatment maximizes the probability $P(\mu_i < 2.5\ mmol/L\ )?$"*

(treatment hierarchy question 2).

This orders the treatments according to their areas below the dotted line in Figure 1 and indicates C as the preferable treatment.

Note that treatment hierarchy questions can be expressed using absolute or relative treatment effects. In Box 1 we explain why this choice is important when more than two treatments are to be compared.

## RANKING METRICS AND THEIR PROPERTIES

All ranking metrics summarize the distribution of treatment effects estimated in NMA, $\mu_i$ or $\delta_{ij}$, and transform them into a set of numbers, one number (metric) for each treatment. Ranking metrics mostly differ in the way they combine the mean and uncertainty in the estimated $\mu_i$ or $\delta_{ij}$. A summary of the ranking metrics is presented in Table 2.

The role of precision in the estimation of $\mu_i$ or $\delta_{ij}$ when calculating ranking metrics is responsible for the disagreement in the resulting hierarchies. In Box 2 we outline the factors that control the precision in the estimation of $\mu_i$ or $\delta_{ij}$. To explore these further, we compare the hierarchy of treatments whose effects are estimated with different levels of precision in the following hypothetical example.

### Hypothetical example

In Figure 2 we present two possible scenarios for the estimated absolute treatment effects where we have four treatments, named P (placebo), A, B and C. Clearly, the active drugs (A, B and C) are better than P, and we now want to create a hierarchy between the active drugs.

**Hierarchy based on the point estimate of the mean relative or absolute treatment effects**

In many applications, it is implicit that the treatment hierarchy question relates to the center of the distribution $M_i$ (or $D_{ij}$ ). Ranking treatments according to $M_i$ answers the question *"Which* *treatment is associated with the smallest estimated mean value on the studied outcome?*". Similarly, ranking according to $D_{ij}$ answers the equivalent question *"Which treatment is associated with the largest estimated mean advantage compared to all other competitors?"*.

This approach considers only the point estimate $M_i$ in each distribution and incorporates uncertainty in the estimation only to the degree that this contributes to the calculation of $M_i$. This can be justified from decision theory (17). In scenario 1 in Figure **2**, the treatment hierarchy is A, B and then C, while in scenario 2 all three active treatments are equivalent.

**Hierarchy based on probability of a treatment producing the best mean outcome value**

The probability $p_{i,BV}$ that treatment *i* has the best value (BV) for a mean outcome at the population level (the smallest value for a harmful outcome or the largest for a beneficial outcome) is mis-interpreted in many reports of NMAs, as the probability for a treatment to be overall the best option (and often denoted as 'P(best)'). Note again the distinction between a treatment having the best mean outcome at the population level (which is not directly observed) and a treatment being the best treatment option (which is a clinical decision based on observed data). In the context of our example of Figure 2, $p_{A,BV}$ is the probability that the mean outcome $\mu_A$ with treatment A is more favorable than the mean outcome with either treatments B, C or placebo. The probability $p_{i,BV}$ is defined as the probability that *i* 'beats' all other competing treatments and can be directly estimated in a Bayesian setting or in a frequentist setting using re-sampling.

The difference between treatment hierarchy questions often lies in how uncertainty is taken into account. Treatment hierarchy question 1 looks for the preferable treatment by comparing *estimates of the* $\mu_i$, $M_i$, where $M_i$ is computed by averaging over the uncertainty in the estimation of $\mu_i$. Treatment hierarchy question 2 aims instead to find the best treatment by comparing $\mu_i$ themselves; because these are unknown, it instead looks for the treatment is associated with the largest $P(\mu_i < 2.5\ mmol/L)$.

Ranking based on $p_{i,BV}$ answers the question *"Which treatment is most likely to have the best (most desirable) mean value on the studied outcome?"*. Then, the hierarchy will be obtained using $p_{i,BV}$, with larger values corresponding to more preferable treatments. However, a high $p_{A,BV}$ value does not suggest that A is preferable under *any* treatment hierarchy question. In particular, there might be a large probability that A also produces the worse mean outcome among all competitors.

In scenario 1 in Figure 2 the hierarchy using $p_{i,BV}$ agrees with that obtained from ranking $M_i$. In scenario 2, $p_{i,BV}$ give a the hierarchy C, B and A: this differs from the hierarchy when ranking $M_i$ because it reflects differences in the precision with which $\mu_A, \mu_B$ and $\mu_C$ are estimated. However, it cannot be described as a 'wrong' or 'misleading' treatment hierarchy.

**Rankograms and cumulative ranking plots**

An extension to $p_{i,BV}$ considers both tails of the distributions of $\mu_i$, by calculating the probability that a treatment is the best, the worst and all positions in between. The probability $p_{i,r}$ is the probability that treatment $i$ will 'beat' exactly $T-r$ treatments; $p_{i,1}$ is the same as $p_{i,BV}$. The cumulative probability $cp_{i,r} = \sum_{k=1}^{r} p_{i,k}$ is the probability that treatment $i$ 'beats' at least $T-r$ treatments or, equivalently, the probability that $i$ is one of the top $r$ treatments. The plots of $p_{i,r}$ and $cp_{i,r}$ (presented in Table 3 for the scenarios in Figure 2) are termed *rankograms* and *cumulative ranking plots*.

Rankograms do not necessarily imply a treatment hierarchy or answer a specific treatment hierarchy question. A summary measure of the probabilities in rankograms is desirable; several possible options are presented below.

**Hierarchy based on the surface under the cumulative ranking curve (SUCRA)**

A numerical summary of the rankograms is provided by SUCRA (18). SUCRA is calculated as the sum of all cumulative rank probabilities up to $T-1$ divided by $T-1$ (Table 1). For a treatment $i$, $SUCRA_i$ measures the extent of certainty that a treatment beats all other competing treatments. It can therefore answer the question *"Which treatment has the largest fraction of competitors that it beats?"*.

SUCRA synthesizes all ranking probabilities in a single number and reflects the overlap between the treatment effect distributions; the larger the overlap, the more similar are the $SUCRA_i$ values. If all treatments have the same $M_i$ but various degrees of uncertainty, then they all have $SUCRA_i$ =50%. Rücker and Schwarzer suggested a transformation of the one-sided p-values that test the differences between the means of the effect distributions as another way to calculate SUCRA, termed P-Score (4).

Table 3 shows SUCRA values for the three scenarios in Figure **2**. The hierarchy obtained by SUCRA in scenario 1 is in agreement with the hierarchy obtained with $p_{i,BV}$. In scenario 2, SUCRA values are very close together for the three active interventions; the effect $\mu_C$ is estimated with a lot of uncertainty and this results in more overlap between the effects of C and P and consequently the SUCRA values of these two interventions are closer together compared to treatment A.

**Hierarchy based on mean or median rank**

To rank treatments, the mean or median rank for treatment $i$ ($MeanR_i$ and $MedianR_i$) can also be used. Mean ranks are transformations of SUCRAs ($MeanR_i = T - (T-1) \times SUCRA_i$) and can be estimated via P-scores. Consequently, $MeanR_i$ answers the same question as SUCRA. Mean and median ranks are however more intuitively associated with the question *"In the distribution of treatment effect ranks, which treatment has the largest mean(median) rank?"*

respectively. Because of their mathematical relation, mean ranks always result in the same treatment hierarchy as SUCRA. Median ranks might be easier to gauge as they are integers, but the presence of many ties might conceal small differences between treatments.

# THE IMPACT OF IMPRECISELY ESTIMATED TREATMENT EFFECTS ON TREATMENT HIERARCHY

As discussed in the previous sections, the various ranking metrics differ in the way they incorporate uncertainty in the estimation of $\mu_i$. Below we explore further the level of agreement between the ranking metrics $M_i, p_{i,\text{BV}}$ and SUCRA.

**Impact of uncertainty on hierarchies obtained by estimated mean effects $M_i$ and SUCRA**

When the differences in precision across estimates of $\mu_i$ are extreme, the hierarchy obtained with SUCRA can conflict with that obtained with $M_i$. Consider the example in Figure **2** scenario 1. According to $M_i$, the hierarchy is treatment A, then B and finally C. We gradually increase the uncertainty around the estimation of $\mu_A$ as shown in the first column of Table **4**. In the extreme scenario where the variance of $\mu_A$ is 20, the hierarchy obtained with SUCRAs is B, C, A. Increased uncertainty in the estimation of $\mu_A$ results in wider overlap with its competitors. When these competitors have point estimates worse than A (as is the case here) then lower precision in $\mu_A$ leads to lower ranks for A. If the competitors P, B and C had point estimates superior to A, then lower precision in $\mu_A$ would lead to higher ranks for A. In general, when the competitors of a random treatment X have $M_i$ worse than $M_X$, then larger imprecision in $\mu_X$ leads to lower ranks for *X*. The opposite occurs when the competitors of X have more favorable point estimates than $M_X$.
As explained before, this disagreement occurs because SUCRAs (and other probabilistic metrics) incorporate the uncertainty in the effect size estimates, while the point estimates of the mean effects $M_i$ do not.

**Impact of uncertainty on the hierarchies obtained by $p_{i,\text{BV}}$ and SUCRA**

To study further the impact of uncertainty on the differences in hierarchy between $p_{i,\text{BV}}$ and SUCRA, we now assume placebo to have the lowest mean value $M_P = -2$ and the other treatments A, B, C to have $M_i$ 1, 1.5 and 2 respectively. We start with all standard deviations of the estimated distributions equal to 1; then we gradually increase the standard deviation of C, $SD_C$, up to 10. With $SD_C = 2$, $p_{i,\text{BV}}$ suggests treatment C is higher in the hierarchy than A while it needs an $SD_C$ as large as 7.5 for SUCRA to indicate C is higher in the hierarchy than A (Figure **3**).

## DISCUSSION

In this article, we introduced the idea of a treatment hierarchy question, and we suggested that the clinical decision-making problem should be clearly defined at the beginning of every comparative effectiveness review. We showed that each ranking metric could, in principle answer a specific treatment hierarchy question, and when used in this context, every ranking metric provides a valid treatment hierarchy for the corresponding question.

Several articles have criticized the use of ranking metrics in published NMAs. Their criticism is misplaced because there is no universally accepted 'gold standard' treatment hierarchy against which the hierarchy obtained by the various metrics is to be evaluated. All apparent limitations of the ranking metrics result from the fact that they transform a complex set of information (e.g. distributions for the mean treatment effect with location and dispersion) into a set of (univariate) numbers. Using theoretical examples, we have shown that

a) SUCRA values depend on the precision of estimation of treatment effects, but they do not consistently under or over-estimate the rank of the treatments whose effects are imprecise; instead, changes in hierarchy very much depend on the rankings of the other treatments.

b) $p_{i,\text{BV}}$ is more sensitive to differences in precision across treatment effect estimates than SUCRA. Among treatments with the same point estimate, it ranks first the treatment with the most imprecise effect.

c) In the presence of large differences in the precision of the estimated effects, all probabilistic ranking metrics may provide hierarchies that disagree with that of the estimated treatment effects.

Setting up the treatment hierarchy question is not trivial and further research is needed to define the spectrum of questions that an NMA can answer. Part of the difficulty in specifying a treatment hierarchy question is in clear use of language. In this article, we offer some ideas about what the question could be, but some stakeholders may be interested in questions that cannot be answered by any of the existing metrics. Decision making is a complex process that considers several efficacy and safety or tolerability outcomes, clinically important differences in the relative treatment effects, the utilities associated with each outcome value and predictions in real-world conditions. These considerations motivate extensions of the existing ranking metrics or formal decision analysis (19–21).

Treatment rankings have also been criticized for not including assessments of the quality of evidence (22); this applies equally to relative treatment effects. In a pairwise meta-analysis, interventions with large and precise effects are not necessarily preferable if the evidence is of low quality. Similarly, the treatments at the top of the hierarchy should not be blindly recommended, without first scrutinizing the confidence in the results. The risk of bias in the included studies, the amount of heterogeneity, the plausibility of the consistency assumption and the threat of publication bias could all limit the credibility of a treatment hierarchy, just as for effect sizes in pairwise meta-analysis. An attempt to produce statements about the credibility of ranking can be found in Salanti et al. (23) and is subject to ongoing research extending the CINeMA framework (Confidence in Network Meta-Analysis) (24). Systematic reviewers should consider the quality of the evidence when translating numerical results (effect sizes or rankings) into practice recommendations and failure to do so should not be perceived as a shortcoming of the ranking metrics employed.

The main challenge that decision-makers and health-care professionals face is to be aware of the advantages and disadvantages of rankings and to be transparent about the methods used. Even when a treatment hierarchy question is clearly defined, and the appropriate metric used, the importance of ranking interventions is not to provide a 'cookbook' for health care decision making. Interpretation of a treatment hierarchy must ideally extend beyond inspection of the values from ranking measures and draw on the totality of the evidence synthesis results. In this spirit, we recommend that every systematic review explicitly defines in its protocol the treatment hierarchy question it aims to answer, chooses an appropriate ranking metric for that question, and interprets the obtained hierarchy after considering the uncertainty in the treatment effects and the quality of the evidence.

*References*

1. Caldwell DM, Ades AE, Higgins JP. Simultaneous comparison of multiple treatments: combining direct and indirect evidence. *BMJ*. 2005;331(1756-1833 (Electronic)):897–900.

2. Salanti G, Ades AE, Ioannidis JP. Graphical methods and numerical summaries for presenting results from multiple-treatment meta-analysis: an overview and tutorial. *J.Clin.Epidemiol.* 2011;64(1878-5921 (Electronic)):163–171.

3. Lu G, Ades AE. Combination of direct and indirect evidence in mixed treatment comparisons. *Stat. Med.* 2004;23(20):3105–3124.

4. Rücker G, Schwarzer G. Ranking treatments in frequentist network meta-analysis works without resampling methods. *BMC Med. Res. Methodol.* 2015;15:58.

5. Salanti G. Indirect and mixed-treatment comparison, network, or multiple-treatments meta-analysis: many names, many benefits, many concerns for the next generation evidence synthesis tool. *Res. Synth. Methods*. 2012;3(2):80–97.

6. Cipriani A, Higgins JPT, Geddes JR, et al. Conceptual and technical challenges in network meta-analysis. *Ann. Intern. Med.* 2013;159(2):130–137.

7. Hutton B, Salanti G, Caldwell DM, et al. The PRISMA extension statement for reporting of systematic reviews incorporating network meta-analyses of health care interventions: checklist and explanations. *Ann. Intern. Med.* 2015;162(11):777–784.

8. Kibret T, Richer D, Beyene J. Bias in identification of the best treatment in a Bayesian network meta-analysis for binary outcome: a simulation study. *Clin. Epidemiol.* 2014;6:451–460.

9. Mills EJ, Kanters S, Thorlund K, et al. The effects of excluding treatments from network meta-analyses: survey. *BMJ*. 2013;347(1756-1833 (Electronic)):f5195.

10. Veroniki AA, Straus SE, Rücker G, et al. Is providing uncertainty intervals in treatment ranking helpful in a network meta-analysis? *J. Clin. Epidemiol.* 2018;100:122–129.

11. Trinquart L, Attiche N, Bafeta A, et al. Uncertainty in Treatment Rankings: Reanalysis of Network Meta-analyses of Randomized Trials. *Ann. Intern. Med.* 2016;164(10):666–673.

12. Wang Z, Carter RE. Ranking of the most effective treatments for cardiovascular disease using SUCRA: Is it as sweet as it appears? *Eur. J. Prev. Cardiol.* 2018;25(8):842–843.

13. Chiocchia V, Nikolakopoulou A, Papakonstantinou T, et al. Agreement between ranking metrics in network meta-analysis: an empirical study. *BMJ Open*. 2020;10(8):e037744.

14. Zarin W, Veroniki AA, Nincic V, et al. Characteristics and knowledge synthesis approach for 456 network meta-analyses: a scoping review. *BMC Med.* 2017;15(1):3.

15. Committee for Human Medicinal Products. Addendum on estimands and sensitivity analysis in clinical trials to the guideline on statistical principles for clinical trials E9(R1). 2017;

16. Piepoli MF, Hoes AW, Agewall S, et al. 2016 European Guidelines on cardiovascular disease prevention in clinical practiceThe Sixth Joint Task Force of the European Society of Cardiology and Other Societies on Cardiovascular Disease Prevention in Clinical Practice (constituted by representatives of 10 societies and by invited experts)Developed with the special contribution of the European Association for Cardiovascular Prevention & Rehabilitation (EACPR). *Eur. Heart J.* 2016;37(29):2315–2381.

17. Berger JO. Statistical Decision Theory and Bayesian Analysis. 2nd ed. New York: Springer-Verlag; 1985 (Accessed August 3, 2020).(https://www.springer.com/gp/book/9780387960982). (Accessed August 3, 2020)

18. Salanti G, Ades AE, Ioannidis JPA. Graphical methods and numerical summaries for presenting results from multiple-treatment meta-analysis: an overview and tutorial. *J. Clin. Epidemiol.* 2011;64(2):163–171.

19. Brignardello-Petersen R, Johnston BC, Jadad AR, et al. Using decision thresholds for ranking treatments in network meta-analysis results in more informative rankings. *J. Clin. Epidemiol.* 2018;98:62–69.

20. Tervonen T, Naci H, van Valkenhoef G, et al. Applying Multiple Criteria Decision Analysis to Comparative Benefit-Risk Assessment: Choosing among Statins in Primary Prevention. *Med. Decis. Mak.* 2015;35(7):859–871.

21. Mavridis D, Porcher R, Nikolakopoulou A, et al. Extensions of the probabilistic ranking metrics of competing treatments in network meta-analysis to reflect clinically important relative differences on many outcomes. *Biom. J.* 2020;62(2):375–385.

22. Mbuagbaw L, Rochwerg B, Jaeschke R, et al. Approaches to interpreting and choosing the best treatments in network meta-analyses. *Syst. Rev.* 2017;6(1):79.

23. Salanti G, Del Giovane C, Chaimani A, et al. Evaluating the quality of evidence from a network meta-analysis. *PLoS One*. 2014;9(7):e99682.

24. Nikolakopoulou A, Higgins JPT, Papakonstantinou T, et al. CINeMA: An approach for assessing confidence in the results of a network meta-analysis. *PLoS Med.* 2020;17(4):e1003082.

Box 1. Framing The Treatment Hierarchy Question Using Absolute And Relative Treatment Effects

Treatment hierarchy questions may be expressed using *absolute effects* (like means and probabilities) or *relative effects* (like mean differences or risk ratios). Any question that can be expressed using relative effects can also be expressed using absolute effects, but the opposite is not always possible. For example, treatment hierarchy question 1 can be expressed as "Which treatment has the largest reduction in estimated mean post-treatment LDL-C compared to treatment A?". This is the same question, since the treatment with the smallest value of $M_i$ must also have the smallest value of $D_{iA}$ (where we interpret $D_{AA} = 0$). However, treatment hierarchy question 2 cannot be expressed in terms of relative effects $\delta_{ij}$.

Treatment hierarchy questions that cannot be expressed in terms of relative effects can result in treatment hierarchies that might seem inappropriate or counterintuitive. For example, in Figure 1, it might appear strange to prefer treatment C to treatment B based on its lower probability of having a mean LDL-C above 2.5 mmol/L (*treatment hierarchy question 2*), when it also has a larger estimated mean LDL-C (*treatment hierarchy question 1*). We can debate whether treatment hierarchy question 2 is of relevance to a particular decision-making context or unsuitable for producing treatment guidelines. But we cannot argue that the obtained hierarchy C, B, A is wrong, because it correctly answers a valid treatment hierarchy question.

In two-arm randomized trials and pairwise meta-analysis, conclusions are in practice based on the (single) relative effect $\delta_{ij}$. Because the uncertainty in the estimation of $\delta_{ij}$ is usually distributed equally around the centre of the distribution $D_{ij}$ (a symmetric distribution), all treatment hierarchy questions usually give the same answer. For example, if $D_{ij} < 0$ indicating that treatment $i$ has lower estimated LDL-C than treatment $j$, then also the probability that treatment $i$ beats treatment $j$ is greater than ½ and hence greater than the probability that treatment $j$ beats treatment $i$.

Box 2. Sources Of Imprecision In The Estimation Of Treatment Effects In Network Meta-Analysis.

Most NMA models provide estimates of relative treatment effects $\delta_{AP}$, $\delta_{BP}$, $\delta_{CP}$ for A, B, C versus a reference treatment, say P. The absolute treatment effects $\mu_A$, $\mu_B$, $\mu_C$ and $\mu_P$ can be directly estimated from the model or be obtained by combining the $\delta$'s with an estimate for $\mu_P$. Each relative or absolute treatment effect is estimated with uncertainty, which depends on the following three factors:

*1. The amount of direct and indirect information for treatment and treatment comparison.* Little information for an absolute effect is available if the treatment features in few studies, if its studies have large sampling error (small sample size, few events or large standard deviations), if it is compared with few other interventions or is part of few closed loops of evidence. Similarly, the relative effect is uncertain if few studies (or small studies) examine the comparison of interest and if the comparison is part of few closed loops.

*2. The heterogeneity in relative treatment effects.* If the relative treatment effects δ are heterogeneous across studies examining the same comparison, the uncertainty in their estimation will be larger. As absolute effects are estimated from relative effects, heterogeneity in δ will result in more uncertainty in the estimation of $\mu$ as well.

*3. Residual incoherence.* Evidence of large incoherence (disagreement between direct and indirect evidence) should prevent researchers from synthesizing the data. Even if a treatment is involved in evidence loops with small or moderate amounts of incoherence, the credibility of the estimated summary effect decreases. However, this produces estimates with less precision and larger credible intervals only when residual incoherence is explicitly modelled within the NMA.

The first two situations outlined above or their combination always result in increased uncertainty about $\delta$ and $\mu$ and this uncertainty plays a major role when estimating treatment hierarchy using ranking metrics.

Table 1 Summary Of Notation For The Observed And Unobserved Quantities In A Network Meta-Analysis That Compares T Competing Treatments

|  | Unobserved quantities (to be estimated) | Observed quantities (estimates) |
|---|---|---|
| **Means** | Parameters $\mu_i, \delta_{ij}$ | Parameters $M_i, D_{ij}$ |
| **Treatment comparisons** | Treatment $i$ beats treatment $j$, $\delta_{ij} < 0$ | Probability that treatment $i$ beats treatment $j$, $P(\delta_{ij} < 0)$ |
| **Treatment rankings** | Treatment $i$ is the best, $\delta_{ij} \leq 0$ for all $j$ | Probability that treatment $i$ is the best, $P(\delta_{ij} \leq 0 \text{ for all } j)$ |
| **Ranking metrics** |  | Any summary of the estimates above or probabilities |

Table 2 Ranking Metrics Used To Obtain A Treatment Hierarchy In Network Meta-Analysis, Their Formulas And The Treatment Hierarchy Questions They Can Answer. Assuming A Harmful Outcome We Consider That A Treatment ***i*** 'Beats' Treatment ***j*** When The True Mean Values Of The Outcome Fulfil $\boldsymbol{\mu_i < \mu_j}$.

| Ranking Metric | Calculated as | Treatment hierarchy question |
|---|---|---|
| Point estimate of the mean relative or absolute treatment effect $M_i, D_{ij}$ | The center of the estimated distribution of the absolute or relative effect | *"Which treatment has the smallest estimated mean value on the studied outcome?"* *"Which treatment has the largest estimated mean advantage compared to all other competitors?"* |
| Probability of a treatment having the best mean outcome value ($p_{i,BV}$) | $P(\mu_i < \mu_j \text{ for all treatments } j \neq i)$ | *"Which treatment is most likely to have the best (most desirable) mean value on the studied outcome?"* |
| Surface under the cumulative ranking curve $SUCRA_i$ | $\dfrac{\sum_{r=1}^{T-1} \sum_{j=1}^{r} p_{i,j}}{T-1}$ | *"Which treatment has the largest fraction of competitors that it beats?"* |
| Mean rank $MeanR_i$ | $\sum_{r=1}^{T} p_{i,r} \times r$ | *"In the distribution of treatment effect ranks, which treatment has the largest mean rank?"* |
| Median rank $MedianR_i$ | The value satisfying $\sum_{r=1}^{medianR_i-1} p_{i,r} \leq \frac{1}{2}$ and $\sum_{r=medianR_i+1}^{T} p_{i,r} \geq \frac{1}{2}$ | *"In the distribution of treatment effect ranks, which treatment has the largest median rank?"* |

$T$: The number of competing treatments, $p_{i,r}$: the probability that treatment I will produce the *r*-th more favorable value (or will 'beat' exactly *r* treatments)

Table 3 Ranking Metrics For The Hypothetical Scenarios Presented In Figure 1.

|  | Ranking metric | P | A | B | C |
|---|---|---|---|---|---|
| Scenario 1 | $p_{i,BV}$(%) | 0.2 | 48 | 31.7 | 20.1 |
|  | $cp_{i,2}$(%) | 1.4 | 79.3 | 67.7 | 51.7 |
|  | $cp_{i,3}$(%) | 8 | 98.8 | 97.5 | 95.7 |
|  | $cp_{i,4}$(%) | 100 | 100 | 100 | 100 |
|  | SUCRA (%) | 3.2 | 75.2 | 65.6 | 56.0 |
|  | P-score | 3.2 | 75.2 | 65.6 | 56.0 |
|  | Mean Rank | 3.9 | 1.7 | 2.0 | 2.3 |
|  | Median Rank | 4 | 2 | 2 | 2 |
| Scenario 2 | $p_{i,BV}$(%) | 0.1 | 26.1 | 33.1 | 40.7 |
|  | $cp_{i,2}$(%) | 0.6 | 73.9 | 66.6 | 58.9 |
|  | $cp_{i,3}$(%) | 7.4 | 99.9 | 98.6 | 94.1 |
|  | $cp_{i,4}$(%) | 100 | 100 | 100 | 100 |
|  | SUCRA (%) | 2.6 | 66.6 | 66.1 | 64.7 |
|  | P-score (%) | 2.6 | 66.6 | 66.1 | 64.7 |
|  | Mean Rank | 3.9 | 2.0 | 2.0 | 2.1 |
|  | Median Rank | 4 | 2 | 2 | 2 |

The probability $p_{i,j}$ is the probability that treatment *i* can produce the *r*-best mean value, $r = 1, 2, 3, 4$ (e.g. the best (BV), second best, etc.) on the studied outcome; $cp_{i,j}$ are the cumulative ranking probabilities. SUCRA: the surface under the cumulative ranking curve.

Table 4 Example Showing The Impact Of Increased Imprecision Associated With The Estimation Of The Mean Outcome With Intervention A. The First Column Shows The Outcome Distribution Of $\mu_A$ As A Normal Distribution With Mean 1 And Standard Deviation That Ranges From 3 To 20. The Distributions For The Other Interventions Are $\mu_B \sim N(2,3)$ $\mu_C \sim N(3,3)$ And $\mu_P \sim N(10,3)$ As Shown In Figure 2, Scenario 1.

| Absolute treatment effect $\mu_A$ | SUCRA (%) | | | |
|---|---|---|---|---|
|  | P | A | B | C |
| $\mu_A \sim N(1,3)$ | 3.2 | 75.3 | 65.6 | 55.9 |
| $\mu_A \sim N(1,10)$ | 9.1 | 63.9 | 67.5 | 59.5 |
| $\mu_A \sim N(1,15)$ | 11.9 | 59.9 | 67.9 | 60.3 |
| $\mu_A \sim N(1,20)$ | 13.6 | 57.7 | 68.1 | 60.6 |

Figure 1 Hypothetical Example Of Three Interventions A, B And C Aiming To Reduce LDL-C Levels. The Distributions Refer To The True Population Mean $\boldsymbol{\mu_i}$ For Post-Treatment LDL-C Levels And Is The Result Of The Synthesis Of Randomised Trials That Compare Pairs Of Drugs. The Vertical Dashed Line Represents The Assumed Population-Level Target Of $\boldsymbol{2.5}$ **mmol/L**.

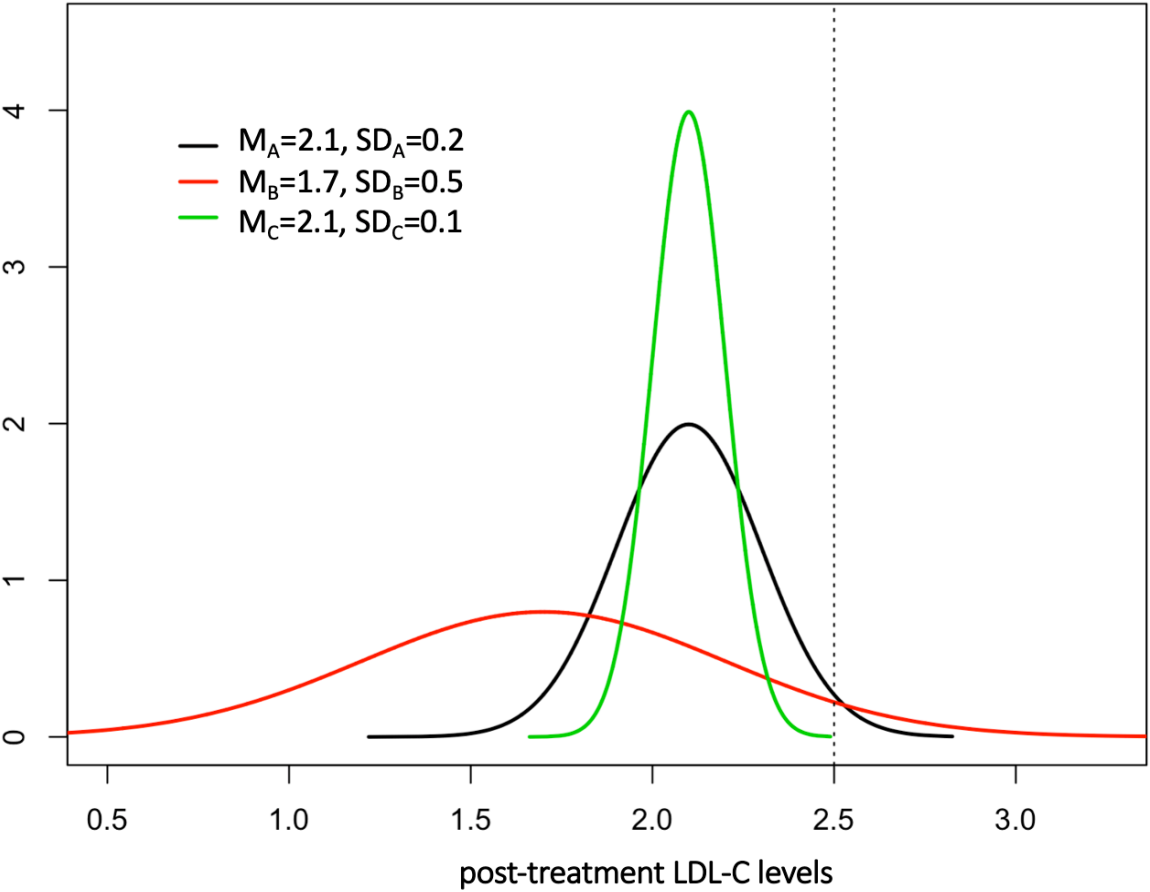

Figure 2 Hypothetical Example: Estimation Of The Absolute Effects, $\boldsymbol{\mu_i}$ Of Three Active Treatments And Placebo With Means $\boldsymbol{M_i}$ And Standard Deviations $\boldsymbol{SD_i}$.

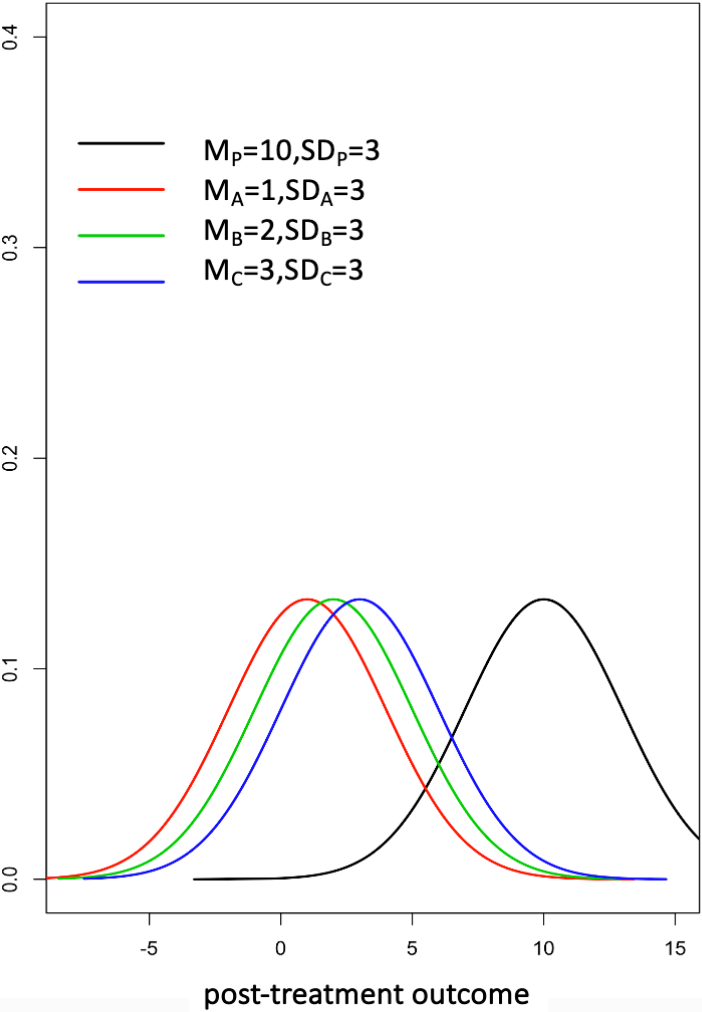
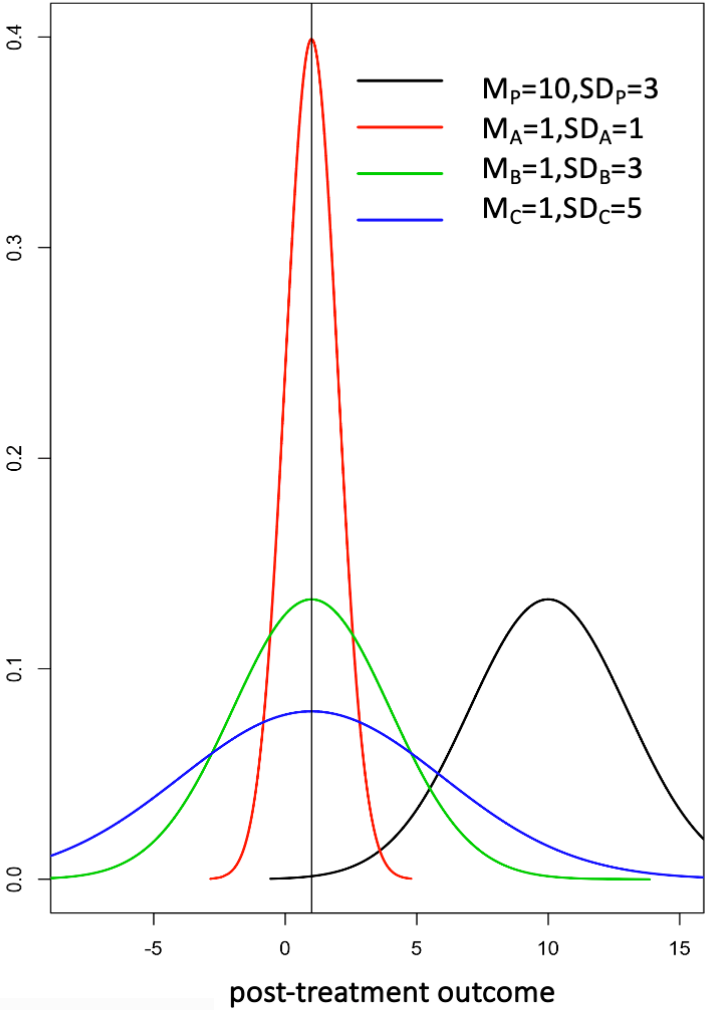

Figure 3 Differences In Hierarchy With SUCRA (Upper Panel) And P(Best Outcome) $p_{i,BV}$ (Lower Panel) When The Standard Deviation For The Absolute Effect Of C, $SD_C$, Increases From 1 To 10. The Other Effects Are $M_C=2$, $M_A=1$, $SD_A=1$, $M_B=1.5$, $SD_B=1$, $M_P=-2$, $SD_P=1$.

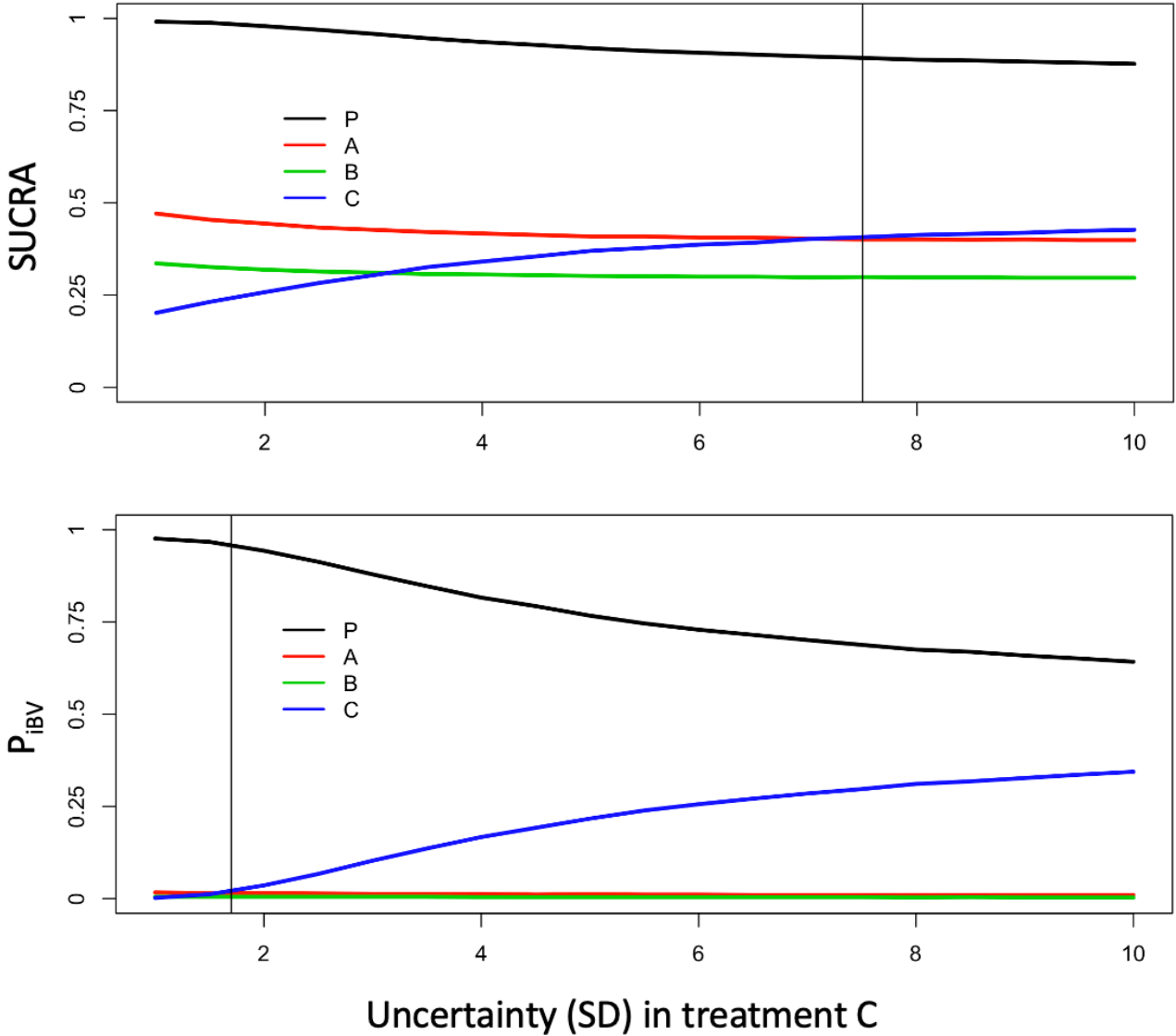